\documentclass[twocolumn,showpacs,preprintnumbers,amsmath,amssymb,floatfix]{revtex4}
\usepackage{graphicx}
\usepackage{dcolumn}
\usepackage{bm}
\begin{document}

\preprint{PRD}

\title[Captures by SMBHs]{Capture Rates of Compact Objects by
  Supermassive Black Holes}

\author{Jos\'e Antonio de Freitas Pacheco and Charline Filloux}
\affiliation{Dpt. CASSIOPEE \\ Observatoire de la C\^ote d'Azur, \\ BP 429
06304 Nice (France). }
\email{pacheco@obs-nice.fr, filloux@obs-nice.fr}

\author{Tania Regimbau}
\affiliation{Dpt. ARTEMIS \\Observatoire de la C\^ote d'Azur, \\ BP 429
06304 Nice (France).}
\email{regimbau@obs-nice.fr}
\date{\today}
\begin{abstract}

Capture rates of compact objects were calculated by using a recent
solution of the Fokker-Planck equation in energy-space, including
two-body resonant effects. The fraction of compact objects (white dwarfs, neutron stars 
and stellar black holes) was estimated as a function of the
luminosity of the galaxy from a new grid of evolutionary
models. Stellar mass densities at the influence radius of central supermassive
black holes were derived from brightness profiles obtained by Hubble Space Telescope observations.
The present study indicates that the capture rates scale as $\propto
M_{bh}^{-1.048}$, consequence of the fact that dwarf galaxies have
denser central regions than luminous objects. If the mass distribution
of supermassive black holes has a lower cutoff at $\sim 1.4\times
10^6$ M$_{\odot}$ (corresponding to the lowest observed supermassive
black hole mass, located in M32), then 9 inspiral events are expected to be seen 
by LISA (7-8 corresponding to white dwarf captures and 1-2 to
neutron star and stellar black hole captures) after one year of operation.
However, if the mass distribution extends down to
$\sim 2\times 10^5$ M$_{\odot}$, then the total number of expected
events increases up to 579 (corresponding to $\sim$ 274 stellar black
hole captures, $\sim$ 194 neutron star captures and $\sim$ 111 white dwarf captures). 

\end{abstract}

\pacs{}

\maketitle

\section{\label{intro}Introduction}

Dynamical evidences suggest that most, if not all elliptical galaxies
and bulges of spirals host supermassive black holes (SMBHs) with
masses ranging from $\sim 10^6$ M$_{\odot}$ up to few $10^9$ M$_{\odot}$
\cite{rich98,koge01}. The most convincing example is the Milky Way, in
which the orbital motion of stars around the radio source SgrA*
indicates the presence of a massive compact object inside a radius
less than 10$^{-3}$ pc, having a mass of $\sim 2-3 \times 10^6$ M$_{\odot}$ \cite{scho02}.  

These SMBHs are generally embedded in dense stellar environments and,
consequently, they may capture stars which either will be promptly
swallowed through the horizon or will inspiral by emitting
gravitational waves. Interactions between stars and SMBHs in dense
galactic nuclei have been the focus of many investigations in the past
years, since these events are potential sources for the 
{\it Laser Interferometer Space Antenna} (LISA). The planned LISA noise
spectral density is minimized around frequencies $\sim$ 3-30 mHz, 
setting the black hole mass interval of interest in the range $\sim 10^5 - 5\times 10^6$ M$_{\odot}$. 
Main sequence stars, which dominate a given stellar population, will
be disrupted by tidal forces if the black hole mass is less than $\sim
2 \times 10^8$ M$_{\odot}$, since the Roche limit will be inside their
gravitational radius. Thus, only the capture of compact objects like
white dwarfs, neutron stars and stellar black holes may produce a
gravitational signal relevant for LISA.

Inside the influence radius $r_{bh}$ of the SMBH, the orbits are dominated by
the gravitational potential of central hole  but are perturbed by nearby
stellar encounters. This means that the total mass under the form of stars
inside $r_{bh}$ is less than the mass of the SMBH and that the radius of this
dynamical sphere of influence can be estimated by equating
the potential energy of the SMBH, $GM_{bh}/r_{bh}$, to the kinetic energy of the 
stars, $\frac{3}{2}\sigma^2_{1D}$ \cite{pee72}. 
In the one hand, stellar-stellar interactions may
produce important variations in energy and angular momentum per
orbital period, leading to a prompt capture of the star if the
resulting angular momentum is less than a critical value $J_{crit} =
4GM_{bh}/c$, for a Schwarzschild black hole. The phase-space volume
defined by $J \leq J_{crit}$ is dubbed the ``loss cone''
\cite{bawo76,bawo77,lisha77} and the infall of these stars in
eccentric orbits onto the SMBH is set by the rate at which relaxation
processes repopulate the loss cone orbits. On the other hand, stars in
tightly bound orbits may undergo a ``diffusion" in J-space with a
small step size $\Delta J$ per orbital period. 
In this case, they lose angular momentum mainly by gravitational
radiation, susceptible to be detected
by LISA. In the absence of the central SMBH, the diffusion-in is
balanced by the diffusion-out of the ``loss cone'' and the net flow is
zero, which is not the case when a secular decay of low angular
momentum orbits occurs due to gravitational radiation.

Due to the complexity of the energy and angular momentum transfer,
predicted capture rates found in the literature may vary by orders of
magnitude. An early Monte Carlo simulation of the stellar random walk in
J-space was performed by \cite{hibe95}, using parameters
appropriate for the core of the dwarf elliptical M32, resulting in a event rate of 
$1.9\times 10^{-8}$ yr$^{-1}$ if compact stars are assumed to represent 10\% of the
total stellar population. Similar simulations were performed by
\cite{frei03} but for parameters appropriate to the galactic
center. In this case, the capture rate of white dwarfs was found to be
$\sim 2-3 \times 10^{-7}$ yr$^{-1}$ and those of neutron stars and
black holes about one order of magnitude lower. Analytical estimates
of the capture rate were performed by \cite{sirees97}, who have found,
in particular for M32, a rate comparable to that obtained by
\cite{hibe95}. Analytical and Monte Carlo methods were adopted by
\cite{hoal05} and from their approach a typical event rate 
of few $\times 10^{-9}$ yr$^{-1}$ per galaxy was derived.

As mentioned above, the capture rate is grossly given by the ratio between the number 
of stars inside the black hole influence radius and the two-body 
relaxation time at this position \cite{bawo76,lisha77}.
The relaxation time is essentially a measure of the timescale required
for uncorrelated two-body interactions to change the original energy
$E$ of a given star by $\Delta E \sim E$. This timescale is also
comparable to that required for the specific angular momentum to change by an amount
comparable to the maximum angular momentum for that energy, corresponding to circular orbits.
In dynamical systems where the potential induces precessing orbits, as
long as the precession timescale (time required to change the argument
of the periapse by $\pi$) is larger than the orbital period, two-body 
interactions last longer and can be imagined as ``correlated". 
The consequences of coherent interactions were recently discussed by
\cite{hoal06}, who have concluded that the steady stellar flow may
increase by almost one order of magnitude in comparison with a
current of stars driven only by non-resonant relaxation.
 
In this work the capture rate of compact objects by SMBHs present in the center 
of early-type (E+S0) and bulges of disky galaxies was estimated from the steady current solution of the
Fokker-Planck equation in energy-space obtained by \cite{hoal06}, 
including resonant and non-resonant interactions. For a sample of E/S0 galaxies,
the stellar density at the influence radius was estimated from the brightness profile 
derived from Hubble Space Telescope (HST) observations, permitting the
derivation of scaling relations either between the capture rate and the total luminosity
of the ``hot" stellar component of the galaxy or the mass of the central SMBH. The present
analysis indicates that dwarf-compact galaxies, besides
hosting SMBHs with masses in the range of interest for LISA, have the
highest capture rates since their cores have stellar densities higher than giant
ellipticals. Convolving with the mass distribution of SMBHs, 
the expected number of events to be detected by LISA after one 
year of operation was estimated. A conservative estimate, including only galaxies 
brighter than M32, indicates that about 9 events are expected to be detected by LISA.
If objects as faint as $M_B = -14$ are supposed to host SMBHs, extending the mass
integration down to $\sim 2\times 10^5$ $M_{\odot}$, the expected number of events
increases dramatically to values around 580.  
This paper is organized as follows: in Section II, the capture rate and the astrophysical parameters
required for its calculation are discussed; in Section III, the maximum redshift at which
an inspiral capture can be detected by LISA is computed as a function of the black hole mass;
in Section IV, using the previous results, estimates of the number of events
expected to be detected by LISA in one year of observations are given and finally, in
Section V, our main conclusions are summarized.
 
\section{The capture rate}

As discussed in \cite{hoal06}, resonant relaxation may increase 
the inspiral rate by almost one order of magnitude relative to rates
predicted by assuming only uncorrelated two-body interactions. This is
a consequence of the fact that symmetries in the gravitational
potential restrict the evolution of the orbits \cite{tre05}, 
increasing the time interval during which stars exert mutual torques. 
In the present work the results obtained by \cite{hoal06} are used to estimate 
the capture rate in elliptical and bulges of spiral galaxies. Notice that the solution obtained
by \cite{hoal06} when applied to the galactic center predicts a capture
rate of white dwarfs of about $1.2\times 10^{-7}$ yr$^{-1}$, in agreement with
the results derived from Monte Carlo simulations by \cite{frei03}.

The capture rate of compact objects $R_{\mathrm{co}}$, including both resonant 
and non-resonant relaxation is given by
\begin{equation}
R_{co} =\frac{8\pi^2}{3\sqrt{2}}\rho^2_{bh}r^3_{bh}\frac{G^2\mathrm{ln}\Lambda}{\sigma^3_{bh}}f_{co}Q(x_{gw})
\label{rate}
\end{equation}
where $\rho_{\mathrm{bh}}=\rho(r_{\mathrm{bh}})$ is the stellar mass density at the influence 
radius $r_{bh}=2GM_{bh}/3\sigma_{bh}^2$,
$\Lambda$ is the ``Coulomb factor" expressing the ratio between the
maximum and the minimum impact parameter of stellar interactions
inside the influence radius, $\sigma_{bh}$ is the line-of-sight
velocity dispersion at the influence radius and $f_{co}$ is the fraction of compact objects of
the stellar population present in central region of the galaxy. $Q(x)$
is the dimensionless net steady rate at which stars flow to energies larger
than $x$. The dimensionless energy is defined by $x =
E/\sigma_{bh}^2$, where $E$ is the binding energy of the star per unit
of mass. Only stars with energies $E \geq E_{\mathrm{gw}}$ can inspiral into
the SMBH without being scattered to a wider orbit of lower energy. The
critical energy $E_{gw}$ can be estimated by equating the inspiral
timescale due to gravitational radiation losses with the timescale required for collisions
produce a variation $\Delta J$ in the angular momentum of the order of $J$,
\cite{hoal05}. In this case one obtains for the dimensionless energy
\begin{equation}
 x_{gw}=2.4\times 10^{-5}\frac{M_6^{3/2}(\rho_{bh}\mathrm{ln}\Lambda)^{2/3}}{\sigma_{200}^4}
\end{equation}
where $M_6$ is the black hole mass in units of $10^6$ M$_{\odot}$,
$\sigma_{200}$ is the line-of-sight velocity dispersion in units of
$200$ km/s and the stellar density at the influence radius is given in
M$_{\odot}$pc$^{-3}$.

The dimensionless flow rate $Q(x)$ is derived from the Fokker-Planck
energy equation including sink terms due to resonant and non-resonant
relaxation. The numerical solution derived by \cite{hoal06} depends on
a factor $\chi$, which characterizes the efficiency of the resonant
relaxation. Here the conservative value $\chi$ = 1 is adopted. In
order to facilitate the numerical computations, the solution derived
by \cite{hoal06} was fitted by a polynomial accurate enough for our
purposes, valid in the range $1 \leq x \leq 550$, e.g.,
\begin{eqnarray}
\mathrm{log} Q(x) = -0.4906-0.4545\mathrm{log}(x)+0.3826\mathrm{log}^2(x)\nonumber\\
-0.2146\mathrm{log}^3(x)
\end{eqnarray} 

\subsection{The fraction of compact stars}

In eq.\ref{rate}, besides the stellar density, the fraction of compact
objects $f_{\mathrm{co}}$ is one of the astrophysical parameters required to
compute the capture rate. For a given stellar population, the fraction
of white dwarfs, neutron stars and black holes depends on {\it the age and
on the initial mass function} (IMF).

If the spectrophotometric indices of elliptical galaxies are
interpreted in terms of single stellar populations, then ages ranging
from few Gyr up to tens Gyr are obtained \cite{tra00,tho05}. Elliptical
galaxies are certainly not single stellar systems in spite of the fact
that the bulk of their stars was formed in a short lapse of time. The
build up of chemical elements requires successive stellar generations,
indicating that those systems are constituted by a population
mix. Multi-population models point to a small age spread around 12
Gyr \cite{imipa03}. A second point to be emphasized, consequence of
these multi-population models, is that the IMF must be flatter
than the usual Salpeter's law in more massive galaxies, in order to explain 
the trend of the $\alpha$-elements abundance ratio with the luminosity 
of the galaxy \cite{imipa03,mato87}. As a consequence, the relative
number of neutron stars and stellar black holes in luminous galaxies
are higher than in dwarf objects.

Here an up-graded version of the grid of models by \cite{imipa03} was used to
estimate the fraction of compact stars in elliptical galaxies of
different masses. Models with an IMF of the form $\xi(m) \propto
m^{1+x}$ were developed and were required to reproduce the observed trends 
in the color-magnitude diagram, (U-V) versus $M_V$, as well as in the
plane of the indices $Mg_2$ and $<Fe>$. This was obtained by adjusting 
free parameters like the star formation efficiency and
the exponent $x$ of the IMF. Stars with masses $m$ lower than 9.0
M$_{\odot}$ give origin to a white dwarf of mass
$m_{wd}=0.116m+0.455$, while progenitors in the range 9-50 M$_{\odot}$ 
were supposed to give origin to a neutron star of mass 1.4 M$_{\odot}$.
Above 50 M$_{\odot}$ stars are supposed to produce a black hole of 10
M$_{\odot}$. The upper limit of $\sim$ 50 M$_{\odot}$ to form a black
hole is consistent with data on X-ray pulsars present 
in binary systems like Wray 977 \cite{ka95} and  Westerlund 1 \cite{muno05}.

\begin{figure}
\centering
\includegraphics[angle=270,width=0.9\columnwidth]{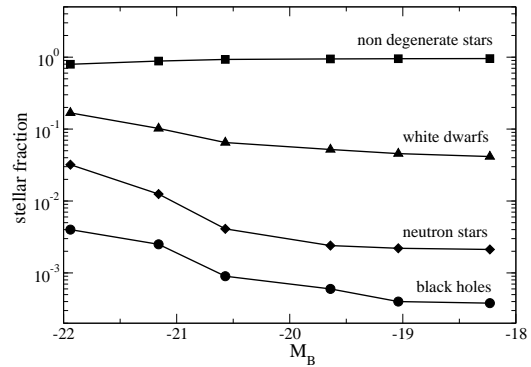}
\caption{Theoretical fractions of non-degenerate stars and of compact objects as
a function of the absolute magnitude of the galaxy}
\label{fraction}
\end{figure}

The expected fraction of non-degenerate stars (essentially main sequence stars), white dwarfs,
neutron stars and black holes as a function of the absolute 
magnitude $M_B$ of the galaxy is shown in fig.\ref{fraction}. In very massive galaxies with $M_B
\approx -22.0$, the fraction of white dwarfs may reach values as high
as 16\%, whereas the fraction of neutron stars and black holes attain 
values of about 3.2\% and 0.4\% respectively.  These results will be used
in our capture rate estimates.

\subsection{The stellar density}

Another important astrophysical quantity required to compute the
inspiral rate from eq.\ref{rate} is the stellar density at the
influence radius. Near the center, the surface brightness along a given axis may be written as
\begin{equation}
I(r) = I_0 + I_1r^2 + {\cal O}(r^4)
\label{intensity1}
\end{equation}
where $r$ is the projected distance from the center. According to
\cite{ritre86}, a galaxy satisfying the above relation can be said to
have an ``analytic core" and its core radius $R_c$ is defined by the
condition $I(R_c)=\frac{1}{2}I(0)$. However, HST data have shown that
real cores are not analytic but display shallow power-law cusps into the resolution 
limit, which brightness can be fitted by the so-called Nuker law \cite{faber97}, e.g.,
\begin{equation}
I(r)= 2^{(\beta-\gamma)/\alpha}I_b(\frac{r_b}{r})^{\gamma}\lbrack 1+
(\frac{r}{r_b})^{\alpha}\rbrack^{(\gamma-\beta)/\alpha}
\label{intensity2}
\end{equation}
where $I_b$, $r_b$, $\alpha$, $\beta$ and $\gamma$ are fitting
parameters. According to \cite{lauer95}, essentially two types of
galaxies can be distinguished on the basis of the aforementioned
parameters: a) ``core'' galaxies, which have an inner logarithmic
slope $\mid d\mathrm{log} I/d\mathrm{log} r\mid < 0.3$ and b) ``power law'' galaxies,
which show a fairly steep brightness profile for $r < r_b$. The former
class  includes luminous objects whereas the later includes fainter
galaxies. These properties are fundamental to understand the stellar
environment in which the central black hole is embedded.

Once the intensity profile is known, the luminosity density $\rho_L$
can be derived by inverting an Abel integral \cite{geb96}, e.g.,
\begin{equation}
\rho_L(R) = -\frac{1}{\pi}\int^{\infty}_R\frac{(dI(r)/dr)}{\sqrt{r^2-R^2}}dr
\label{density}
\end{equation}
and the stellar mass density can be obtained by multiplying the luminosity
density by the mass-to-luminosity ratio $\Upsilon_V = {\cal M}/L_V$ in solar units. 
From our models,
\begin{equation}
\mathrm{log} \Upsilon_V = -1.148-0.0956M_V
\label{mlr}
\end{equation} 

The stellar density at the influence radius was computed numerically using 
eqs.\ref{intensity2}, \ref{density} and \ref{mlr}. The parameters defining the
brightness profile as well as absolute magnitudes were taken from
reference \cite{faber97}, corrected for a Hubble constant $H_0 = 70$ kms$^{-1}$Mpc$^{-1}$, adopted in the present work. Black hole masses
required to compute the influence radius, when available were taken 
from \cite{mefe01a,tre02} or computed from the relation
\begin{equation}
M_{bh} = 1.4\times 10^8\sigma_{200}^{4.26}\, \mathrm{M}_{\odot}
\end{equation}
which is slightly different from the original fits derived by
\cite{mefe01b} or \cite{tre02}, since here only E and S0 galaxies were 
included (bulges of spirals were excluded from the fit).

\begin{figure}
\centering
\includegraphics[angle=270,width=0.9\columnwidth]{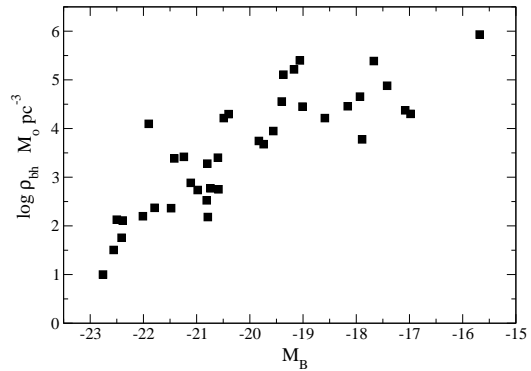}
\caption{Stellar densities at the influence radius as a function of the absolute B-magnitude
for galaxies included in the sample by Faber et al. \cite{faber97}.
Absolute magnitudes were taken from LEDA database \cite{leda}.}
\label{dens}
\end{figure}

Fig.\ref{dens} shows the stellar density at the influence radius as a function of the 
absolute magnitude of the galaxy. This plot just confirms the known fact that 
dwarf ellipticals have central densities considerably higher than bright E-galaxies, which 
is essentially a consequence of ``power-law" profiles present in the former objects and
the existence of a fundamental plane, correlating structural
parameters like the central velocity dispersion, the effective
brightness and the effective radius. As a corollary, SMHBs with masses
around $\sim 10^{5-6}$ M$_{\odot}$ are always embedded in denser stellar environments, thus
having higher capture rates, than those
surrounding SMBHs with masses around $10^{8-9}$ M$_{\odot}$, which have capture rates one or two
orders of magnitude lower.

\subsection{The estimated inspiral rates}

For galaxies included in the sample studied by \cite{faber97}, capture rates were calculated
from eq.\ref{rate}, using the fraction of compact objects, stellar density and influence radius
estimated according to the procedure described above. The resulting capture rate of white dwarfs
scales with the luminosity of the galaxy as
\begin{equation}
R_{wd} = \frac{8.31\times 10^4}{(L_B/L_{B,\odot})^{1.31}} \,\,\mathrm{yr^{-1}}
\label{capture_l}
\end{equation}
whereas the scaling with the SMBH mass is
\begin{equation}
R_{wd} = \frac{0.86}{(M_{bh}/\mathrm{M}_{\odot})^{1.048}} \,\,\mathrm{yr^{-1}}
\label{capture_m}
\end{equation}
The capture rate of neutron stars and stellar black holes can be obtained
by multiplying the equations. above respectively by $0.0511$ and $0.00916$.
In spite of these small rates, the gravitational signal resulting from
the capture either of a neutron star or a stellar black hole can be detected much further
away than that resulting from the capture of a white dwarf, which makes their
detection rate comparable, as we will demonstrate later. Notice that due to the square
dependence on the stellar mass density in eq.\ref{rate}, the capture rate {\it decreases}
with increasing SMBH masses, since more massive holes live in less dense stellar environments.

If one assumes that the relaxation timescale $t_R$  is comparable to the
mean population age, e.g., 12 Gyr, then the stellar density and the velocity
dispersion at $r_{bh}$ must satisfy
\begin{equation}
\rho \approx 0.154\sigma_{1D}^{2.905} \,\, \mathrm{M}_{\odot}\mathrm{pc}^{-3}
\end{equation}
where $\sigma_{1D}$ is given in km/s. On the other hand, densities at the influence
radius derived in the previous section and the line-of-sight velocity dispersion are
approximately related as 
\begin{equation}
\rho \approx 1.44\times 10^{15}\sigma_{1D}^{-5.137} \,\, \mathrm{M}_{\odot}\mathrm{pc}^{-3}
\end{equation}   
From these two relations, in order that $t_R \leq$ 12 Gyr, one must have
$\sigma_{1D} \leq$ 100 km/s and/or $\rho_{bh} \geq 9.0\times 10^4$ M$_{\odot}$pc$^{-3}$. This
means that only galaxies {\it fainter} than $M_B = -18$ have relaxation timescales
less or equal 12 Gyr and can satisfy the steady conditions required by eq.\ref{rate}. Galaxies
brighter than this limit are not able to refill their ``loss cones" within a Hubble time and
probably have a non steady inspiral flow. Therefore, we will restrict the application of eq. \ref{capture_l}
or eq. \ref{capture_m} to objects fainter than the aforementioned limit, which corresponds to galaxies
hosting central black holes with masses around $\sim 8\times 10^6$ M$_{\odot}$.

Eqs. \ref{capture_l}, \ref{capture_m} are valid, in principle, only for early-type galaxies. However, here
they will be also applied to bulges. Bulges constitute, in general, an extension of
low luminosity ellipticals. They follow quite well the $Mg_2-\sigma_{1D}$  diagram
of ellipticals \cite{tapaco96}, the SMBHs present in the center of bulges follow approximately the same
$M_{bh}-\sigma_{1D}$ relation obeyed by early-type galaxies (but having probably a slightly steeper slope
although more data are needed to confirm such a trend \cite{val05}) and stellar densities 
derived at the influence radius, at least for two cases, are in agreement with the results 
displayed in fig.\ref{dens}. These are the bulges of the Milky Way, if we use the 
results by \cite{gen03} and that of NGC 1316,
if we use the procedure described above and data from \cite{faber97}.

\section{The redshift space probed by LISA}

Estimates of the number of events expected to be detected by LISA demand
a previous evaluation of the volume of the universe probed by the detector
or, in other words, an evaluation of the maximum redshift at which a given 
inspiral gravitational signal can be seen.

For the inspiral up to the last stable orbit (LSO), theoretical templates of
the waveform $h(t)$ will probably be available and the method of matched filtering
can be used. In this case, the resulting signal-to-noise ratio (SNR) is \cite{fla98}
\begin{equation}
(\frac{S}{N})^2=4 \int_0^{\infty} d\nu \frac{\vert (\tilde{h}_+(\nu)F_+(\nu) + 
\tilde{h}_{\times}(\nu)F_{\times}(\nu))\vert^2}{S_n(\nu)}
\label{snr}
\end{equation}
where $S_n(\nu)$ is the spectral density of the strain noise of the detector and 
the Fourier transform of the two polarization amplitudes,
$\tilde h_+$ and $\tilde h_{\times}$, depends on the source orientation, on the detector beam
pattern functions $F_+$ and $F_{\times}$, on the polarization and on the source
sky location.
Averaging over source positions and orientations one obtains \cite{cor04}: 
\begin{equation}
(\frac{S}{N})_{\mathrm{rms}}^2=2 \int_0^{\infty} d\nu \frac{S_h(\nu)R(\nu)}{S_n(\nu)}
\label{snr_rms} 
\end{equation}
where the response function of the detector averaged over sky directions and polarizations 
can be expressed as \cite{cor04}:   
\begin{equation}
R(\nu)=\int \frac{d\Omega}{4 \pi}\sum_{A=+,\times}F^{A*}(\hat{\Omega},\nu)F^{A}(\hat{\Omega},\nu)
\label{R} 
\end{equation}
On the other hand, the planned LISA sensitivity is usually given in terms of the quantity 
\begin{equation}
\mid\tilde{h}_{eff}(\nu)\mid^2 = (\frac{S_n(\nu)}{R(\nu)})
\label{heff} 
\end{equation}

The spectral density of the gravitational signal is given by \cite{fla98}
\begin{equation}
S_h(\nu)=\frac{1}{2}\frac{G}{\pi^2c^3}\frac{(1+z)^2}{d_L^2}\frac{1}{\nu^2}\frac{dE}{d\nu_e}((1+z)\nu)
\label{Sh} 
\end{equation}
where $\nu$ and $\nu_e$ are respectively the gravitational wave frequencies at the observer's frame and
at the source related by $\nu_e = (1+z)\nu$,
$d_L=(1+z)r(z)$ is the distance luminosity with $r(z)$ being the proper distance defined by
\begin{equation}
r(z)=\frac{c}{H_0}\int^z_0\frac{dz'}{E(\Omega_i,z')}
\end{equation}
with 
\begin{equation}
E(\Omega_i,z)=\sqrt{\Omega_v + \Omega_m(1+z)^3}
\end{equation}
where $\Omega_v$ = 0.7 and $\Omega_m$ = 0.3 are respectively the present values of the
density parameters due to vacuum and matter (baryonic + non-baryonic) adopted in the
present work, in agreement with the ``concordance" model (\cite{spe03}). 
In eq.\ref{Sh}, the spectral 
energy density of the source in the quadrupole
approximation, which is adequate to estimate the SNR obtained from matched filtering 
\begin{equation}
\frac{dE}{d\nu_e}=\frac{(G\pi)^{2/3}}{3}\mu M^{2/3}\nu_e^{-1/3}
\label{dEdnu} 
\end{equation} 
where the reduced mass $\mu \approx m_{co}$ and the total mass $M \approx M_{bh}$.

Combining eqs.\ref{Sh}, \ref{dEdnu} and \ref{heff} into \ref{snr_rms}, one
obtains
\begin{equation}
(\frac{S}{N})_{\mathrm{rms}}^2=
  \frac{1}{3\pi^{4/3}}\frac{Gm_{co}}{c^3}
  \frac{(GM_{bh})^{2/3}}{r^2(z)(1+z)^{1/3}}
  {\cal I}_{\nu}
\label{snr_final} 
\end{equation}
where 
\begin{equation}
{\cal I}_{\nu} = \int_{\nu_{min}/(1+z)}^{\nu_{max}/(1+z)}\frac{\nu^{-7/3}}{\mid\tilde{h}_{eff}(\nu)\mid^2}d\nu
\end{equation}
The upper limit of the above integral corresponds to the
frequency associated to the LSO, corrected for the observer's frame, e.g.,
\begin{equation}
\nu_{max}= \frac{c^3}{\sqrt{216}\pi GM_{bh}}=\frac{0.004384}{M_6}\,\,Hz
\label{numax} 
\end{equation}
For the lower limit we take the frequency corresponding to about one year
before arrival at the LSO, namely,
\begin{equation}
\nu_{min}=\frac{\nu_{max}}{(1+\frac{0.25m_{co}}{M_6^2})^{3/8}}
\label{numin} 
\end{equation}
where $m_{co}$ is in solar units and $M_6$ is in units of $10^6\,\,M_{\odot}$.

\begin{figure}
\centering
\includegraphics[angle=270,width=0.9\columnwidth]{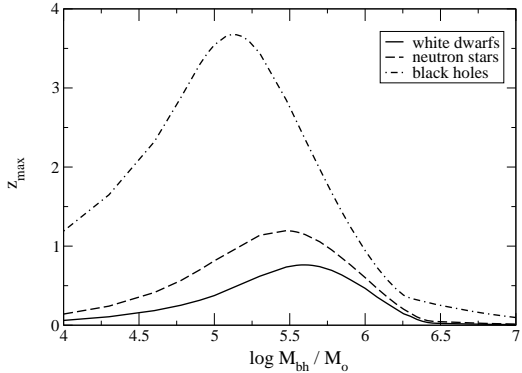}
\caption{Maximum redshift ($z_{max}$) at which a gravitational signal
resulting from an inspiral capture can be detected by LISA operating as
a Michelson interferometer. A signal-to-noise ratio equal to 5 was adopted
and the WD-WD galactic binary confusion noise was included.}
\label{zmax}
\end{figure}

Following the convention adopted in the LISA community, the detectability threshold was
settled to be $S/N = 5$ and from the numerical solution of eq.\ref{snr_final} one obtains, for
a given compact object of mass $m_{co}$, the maximum redshift as function of the SMBH mass $M_{bh}$.  
The effective noise budget sensitivity (see eq.\ref{heff}) used in the calculations refers to a
standard Michelson configuration \cite{ben98}, including the contribution of the galactic binary WD-WD
confusion noise \cite{hbw90,lar00}. The 
resulting curves for white dwarfs ($m_{wd} = 0.7$ M$_{\odot}$), neutron 
stars ($m_{ns} = 1.4$ M$_{\odot}$) and stellar
black holes ($m_{bh} = 10$ M$_{\odot}$) are shown in fig.\ref{zmax}. The farthest inspiral
signal resulting from the capture of a white dwarf corresponds to a redshift $z \sim 0.76$ and
a SMHB mass of about $3.9\times 10^5$ M$_{\odot}$, that resulting from the capture of
a neutron star corresponds to $z \sim 1.19$ and a SMBH of $3.1\times
10^5$ M$_{\odot}$ whereas
the signal resulting from the capture of a stellar black hole by a SMBH of mass
$1.3\times 10^5$ M$_{\odot}$ can be seen up to $z \sim 3.67$.

\section{The expected number of events}

Once the maximum redshift $z_{max}$ at which a given inspiral signal can be 
detected is calculated, the expected number of events $\Gamma_{co}$ in a time interval $T$ can be 
estimated from the equation
\begin{equation}
\Gamma_{co} = T\int_{M_1}^{M_2}\frac{R_{co}(M_{bh})}{(1+z)}
\frac{d{\cal N}(M_{bh},z)}{dM_{bh}}dM_{bh}\int_0^{z(M_{bh})}\frac{dV}{dz}dz
\label{eventnumber}
\end{equation}
In this equation, $R_{co}(M_{bh})$ is the capture rate of a given compact object as a function of the 
black hole mass, as calculated in Section II.B and the term (1+z) in the denominator corrects
for time dilation as the universe expands. Evolutionary effects which may affect the
mass of the central black hole and/or the nearby stellar distribution were neglected
in the present approach but are under investigation. For the SMBH mass distribution, the results 
by \cite{ari02} were used. These authors derived for the local mass distribution function
\begin{equation}
\frac{d{\cal N}(M_{bh},z)}{dM_{bh}} = \frac{\theta_*h^3}{M_0}(\frac{M_0}{M_{bh}})^{1-\gamma}
e^{-(M_{bh}/M_0)^{\epsilon}}(1+z)^3 
\end{equation}
where the parameters $\theta_*$, $M_0$ and $\gamma$ where calculated for the different
morphological types, using the luminosity distribution functions by \cite{mar94}.
The parameter $\epsilon = 0.8\pm 0.1$ depends on the slopes of the adopted
$M_{bh}-\sigma_{1D}$ and $L-\sigma_{1D}$ relations. Since the luminosity function concerns
the ``total" galaxy luminosity, in the case of bulges of spirals the derivation of the black hole
mass distribution depends on the relative bulge-to-total luminosity ratio for the different
morphological types. Since average values for the various morphological types have a
considerable dispersion, this introduces an important source of uncertainty in the parameters
corresponding to disky galaxies. In above 
equation, the conservation of the number of galaxies within the comoving volume was 
included but no other evolutionary effect.
The comoving volume in eq. \ref{eventnumber} is given by
\begin{equation}
\frac{dV}{dz}dz = 4\pi r^2(z)\frac{c}{H_0}\frac{dz}{E(\Omega_i,z)}
\end{equation}

In the second integral of eq.\ref{eventnumber}, the upper limit $M_2$ corresponds to the 
critical luminosity beyond which the relaxation time is larger than 12 Gyr and, consequently, a steady 
stellar current can not be established. The conversion of luminosities into SMBH masses was
performed through the relation 
\begin{equation}
\mathrm{log} M_{bh}/M_{\odot} = -0.835-0.439M_B 
\label{massluminosity}
\end{equation}
If, as we shall see, our results are practically independent on the assumed upper limit $M_2$,
the situation is quite different concerning the lower limit $M_1$, since the capture rate increases
for these less massive black holes. The lower mass cutoff is uncertain since up today observational
searches for intermediate-mass black holes have been discouraging. Presently, the
smallest known central black hole is located in the dwarf elliptical M32, having a mass $\sim 1.4\times
10^6$ M$_{\odot}$ (\cite{mefe01a}). A conservative estimate of the number of events 
can be obtained if this value is taken as the minimum mass. From an observational point of view,
the existence of SMBHs with masses lower than the above limit is still controversial. Analyses of
dynamical data on the dwarf galaxies M33 \cite{me01, geb01} and NGC 205 \cite{val05} lead to
robust upper limits respectively of $\sim 3\times 10^3$ M$_{\odot}$
and $\sim 4\times 10^4$ M$_{\odot}$
to the putative black holes living in the center of these objects. Moreover, searches for
intermediate mass black holes in dwarf-spheroidal satellites of the Milky Way have also been
negative \cite{maca05}. However, these small galaxies were captured by our Galaxy and, in 
the process, collective effects may transfer energy to their central black holes enough to move them far away from
the core, escaping from detection. If direct dynamical evidences of SMBHs in such a mass range are missing,
indirect signals of their presence in narrow-line Seyfert-1 nuclei exist. Using the line width-luminosity-mass
scaling relation established for broad-line AGNs, SMBH masses were estimated by \cite{greho04} for
a sample of 19 galaxies, all in the range $8\times 10^4$ up to
$8\times 10^6$ M$_{\odot}$. 
Therefore, an ``optimistic" estimate was also performed by assuming that
central black black holes are present at least in compact dwarf galaxies and bulges brighter than
$M_B = -14$, since their cores follow quite well the $M_{bh}$-luminosity relation \cite{wewi06}. 
Using eq. \ref{massluminosity}, this luminosity corresponds to a SMBH of about
$2\times 10^5$ M$_{\odot}$.  Such a lower limit were used for  ``optimistic"
estimates of the capture rate of neutron stars and stellar black holes, since for white dwarfs
the adopted lower limit was slightly higher $\sim 6\times 10^5$ M$_{\odot}$, corresponding
to the mass for which the tidal radius is comparable to the gravitational radius. 
The results of our computations indicate that in the {\it conservative} case, 9 inspiral 
events are expected to be detected by LISA. In this case, the estimated probability to have a white dwarf
capture is 84.5\% and those for the capture of a neutron star or a stellar black hole are
8.0\% and 7.5\% respectively. Early-type galaxies are expected to contribute to about
53.8\% of the total number of events, Sa+Sb bulges contribute to about
26.9\% and Sc bulges contribute to about 19.3\%.

If the lower mass integration limit is reduced down to $2\times 10^5$ M$_{\odot}$, there is a dramatic
increase in the expected number of events. 
This occurs because the capture rate increases for
lower SMBH masses as well as the volume of the universe probed by LISA
(see fig.\ref{zmax}). In
this ``optimistic" case, the expected number of events is 579. The contribution of the different
compact objects is now rather different: stellar black holes are expected to represent
about 47.4\% of the total number of events, neutron stars will contribute to about 33.5\%
and white dwarfs, to about 19.1\%. The contribution of the different morphological types
will be practically the same as before, e.g., E+S0 will contribute to about 53.2\%, Sa+Sb, to about
20.5\% and Sc to about 26.3\%.

\section{Conclusions}

Diffusion in energy-space of stars under the gravitational influence of a supermassive
black hole was recently considered in \cite{hoal06}, including phenomenologically
a term in the Fokker-Planck equation, corresponding to two-body resonant relaxation.
These results were used to estimate the capture rate of compact objects in a
sample of early-type galaxies.

From a new grid of evolutionary models for early-type galaxies, including a variable
initial mass function, the fraction of white dwarfs, neutron stars and stellar
black holes was derived as a function of the luminosity (mass) of the galaxy, as
well as the (stellar) mass-to-luminosity ratio, quantities necessary to compute
capture rates. The later was used to estimate stellar mass densities by
inversion of brightness profiles of a sample of early-type objects, obtained from 
HST observations. The analysis of these results indicates that the capture rate
decreases for high luminosity galaxies, which harbor the most massive black holes.
This is a consequence that luminous E-galaxies have cored density profiles whereas
fainter objects have in general ``power-law" profiles, providing a denser stellar
environment to their hosts. 

Interpretation of colors and metallicity indices of early-type galaxies by
evolutionary models indicates mean population ages of about 12 Gyr with a dispersion
of about 2 Gyr. If dynamical ages are comparable, then only systems fainter than
$M_B = -18$ are able to refill their loss cones and maintain a steady stellar
current. Such a luminosity corresponds to a SMBH mass of $\sim 8\times
10^6$ M$_{\odot}$
and constitutes a limit for the application of steady solutions of the Fokker-Planck equation.

Extending our results to bulges of spirals, face their similarity to E-galaxies, the
expected number of events to be detected by LISA in one year of operation was estimated.
Two situations were considered: in the first, a conservative estimate was performed by
restricting the integration over masses in the range defined by the aforementioned upper limit
and the lower limit defined by the lowest SMBH mass detected up today,
e.g., the one located in M32. A second and a more optimistic estimate, extends the lower
mass limit down to $2\times 10^5$ M$_{\odot}$ corresponding to a luminosity of
$M_B = -14.0$. The core of these compact dwarf galaxies, able to host those
undetected up to now black holes, follow the $M-\sigma$ relation \cite{wewi06}
defined by brighter galaxies and bulges, which is encouraging. In this case,
the resulting expected number of events is 579. The inclusion of SMBHs with masses
lower than few $10^6$ M$_{\odot}$ increases not only the capture rate but also
increases the volume ``seen" by the detector (see fig.\ref{zmax}).
In the conservative case, between 7-8 events, probably associated to the
capture of white dwarfs and 1-2 events, probably associated to the capture
of neutron stars and stellar black holes are expected to be seen by LISA after one year of
operation. Reducing the lower mass limit, events
associated to the capture of stellar black holes will be dominant ($\sim$ 274),
followed by those associated to the capture of neutron stars ($\sim$ 194) and
white dwarfs ($\sim$ 111). These aspects associated with the number of events
and the distribution of the different captured compact objects reveal the
potential of gravitational waves as tool for astrophysics, in particular for
scenarios of SMBH formation. 

However, it should be emphasized that the present estimates neglect possible evolutionary
effects. If mean population ages of early-type galaxies and bulges are $\sim$ 12 Gyr, this
means that they were already assembled around $z \sim$ 3.6, which corresponds practically
to the maximum redshift that an inspiral gravitational signal produced by the capture
of a stellar black hole by a SMBH of mass $\sim 1.3\times 10^5$ M$_{\odot}$ can be
detected by LISA with a SNR = 5. No important variations are expected to occur in the
central region of the galaxy as a consequence of the capture of compact objects in 
such a timescale. However, non-degenerate stars will diffuse-in the influence radius
and will be destroyed by tidal forces, representing a mass consumption in a timescale of 12 Gyr
which may be comparable to the that of a $\sim 2\times 10^5$ M$_{\odot}$  black 
hole. This process will modify slowly the central density profile and the SMBH mass. Moreover, hosting 
dark halos are continuously accreting mass \cite{pei04}, affecting the mass history of central
black hole and the structure of the galaxy as well. These problems are presently
under investigation through cosmological numerical simulations and will be reported
in a future paper.

\end{document}